\documentclass[aps, amsfonts, longbibliography, reprint]{revtex4-1}
\usepackage{graphicx}
\usepackage{amsmath}
\begin{document}
\newcommand{\figref}[1]{Fig.~\ref{fig:#1}}
\newcommand{\Figref}[1]{Figure \ref{fig:#1}}
\newcommand{\mysection}[1]{\textit{#1.---}}
\newcommand{\deffig}[4]{%
  \begin{figure}[htb]
    \center
    \includegraphics[width=#2\textwidth]{#3}
    \caption{#4}
    \label{fig:#1}
  \end{figure}
}
\title{
  Universal spectrum structure at nonequilibrium critical points in the (1+1)-dimensional directed percolation
}
\author{Kenji Harada}
\affiliation{Graduate School of Informatics, Kyoto University, Kyoto 606-8501, Japan}
\begin{abstract}
  Using a tensor renormalization group method with oblique projectors
  for an anisotropic tensor network, we confirm that the rescaled
  spectrum of transfer matrices at nonequilibrium critical points in
  the (1+1)-dimensional directed percolation, a canonical model of
  nonequilibrium critical phenomena, is scale-invariant and its
  structure is universal.
\end{abstract}
\maketitle
%
%
\mysection{Introduction}
The universality of critical phenomena has been found not only in
equilibrium systems but also in nonequilibrium ones. The universality
of nonequilibrium critical phenomena in the directed percolation (DP)
has been extensively studied. There is a preferred direction for the
percolation of active objects in the DP. If we regard a preferred
direction as a time, the DP is a type of reaction-diffusion
process. Since the system cannot escape from a state with no active
object, called the absorbing state, the DP process is
nonequilibrium. The critical phase transition of the DP process
between the absorbing state and the other is universal. Experimental
systems of turbulence\cite{Takeuchi:2007co, Takeuchi:2009ia,
  Sano:2016kh} and various theoretical models of reaction-diffusion
processes(See a review \cite{Henkel:2008vv}) belong to the DP
universality class.

In the case of equilibrium critical systems, the renormalization
group(RG) method\cite{Wilson:1971fb, *Wilson:1971bg} is powerful and
conceptually important to understand the universality(See textbooks
\cite{cardy_1996, *goldenfeld2018lectures}). However, the RG approach
for nonequilibrium systems is still challenging. We introduce a tensor
network(TN) representation to attack this problem for the DP.  In this
study, we report a new universal property of DP in a TN
representation. Introducing a new tensor RG (TRG) method for the TN of
dynamical process, we numerically calculate the renormalized critical
tensors of the (1+1)-dimensional DP. We find the universal spectrum
structure of renormalized critical tensors, which is similar to the
conformal tower in the spectrum of critical tensors of the
(1+1)-dimensional equilibrium critical systems\cite{Gu:2009}. In the
following, after we briefly introduce a model of DP and the TN
representation, we will explain our TRG method, and we will report the
spectrum structure of renormalized critical tensors.

%
%
\mysection{Tensor network representation of the (1+1)-dimensional DP}
Domany and Kinzel (DK)\cite{Domany:1984fh} proposed a stochastic
cellular automaton on a square lattice rotated by 45 degrees in
\figref{OPTRG_DP}(a) as the (1+1)-dimensional DP. A site can be either
active or inactive. For the bond DP case, an active state can
percolate to the nearest neighbor sites with a probability $p$ only to
the downward direction. We can regard the row of sites as a
one-dimensional system at a time. The DK cellular automaton is equal
to a one-dimensional Markov process. In the DK cellular automaton, a
probability of an active state depends on the number of active states,
$n$, in the nearest neighbor sites at the previous time as $P[n]$. If
there is no active nearest neighbor just before the current time, the
probability of having an active state is zero, i.e., $P[0]=0$. Thus, a
state with zero density of active objects in a system is an absorbing
state. $P[1]$ and $P[2]$ are free parameters. For the percolation
probability $p$, the bond DP case is defined as $P[1]=p, P[2]=p(2-p)$,
and the site DP case is defined as $P[1]=P[2]=p$. In general, when
$P[1]$ and $P[2]$ increase, the DK cellular automaton displays a
continuous phase transition of the nonequilibrium steady state from
the inactive phase to the active phase in which densities of active
sites are zero and finite, respectively. This phase transition is
critical and universal, which belongs to the DP universality class
except for $P[2]=1$. Since the system of $P[2]=1$ cannot also escape
from the fully active state, there are two absorbing states for
$P[2]=1$. Then, the universal class is different from the other cases
($P[2] \ne 1$), which is called the compact DP. The critical point at
$(P[1], P[2]) = (1/2, 1)$ is also the zero temperature limit of the
one-dimensional kinetic Ising model\cite{Domany:1984fh}.

\deffig{OPTRG_DP}{0.48}{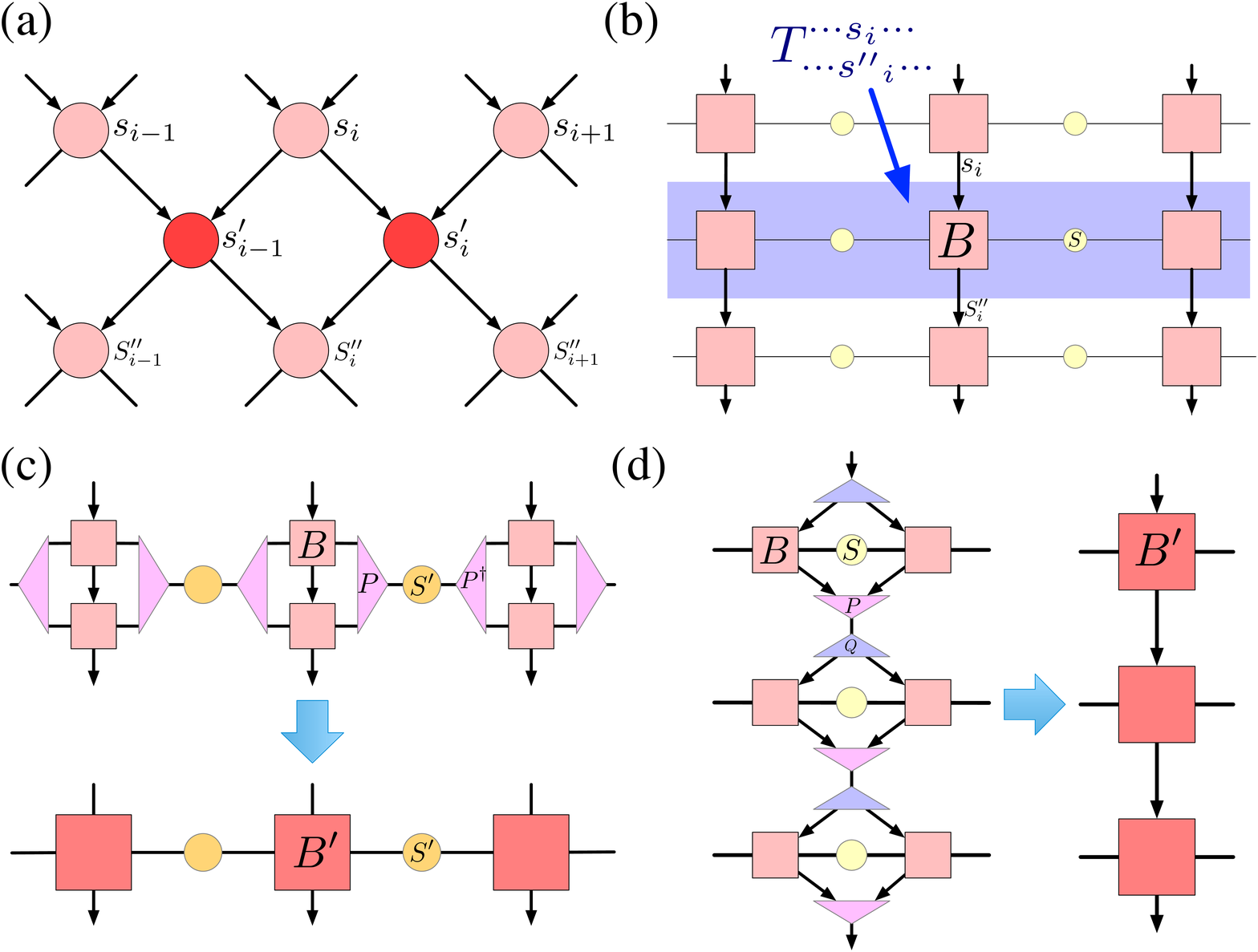}{(a) A lattice of the
  (1+1)-dimensional DK cellular automaton. The horizontal and vertical
  directions denote a spatial and a time axis, respectively. Time
  evolves downward. (b) TN representation of
  $T^{s_1 \dots s_N}_{s''_1 \dots s''_N}$ in the (purple) shaded
  box. (c) Renormalization along a time direction by inserting
  isometries with a new sign diagonal matrix in a spatial direction.
  (d) Renormalization along a spatial direction by inserting oblique
  projectors in a time direction. }

In a one-dimensional Markov process with finite local states, the
state probability of $\{s_i\}$ is written as a rank $N$ tensor,
$P_{s_1 \dots s_N}$, where $s_i$ is a local state variable on a site
$i$ and $N$ is the number of sites. The transfer probability from a
state configuration $\{s_i\}$ to $\{s'_i\}$ is also a tensor as
$T^{s_1 \dots s_N}_{s'_1 \dots s'_N}$. Based on a master equation,
the state-probability distribution at the next time is a tensor
contraction between $P$ and $T$ as
$P_{s'_1 \dots s'_N} = \sum_{s_1 \dots s_N} T^{s_1 \dots
  s_N}_{s'_1 \dots s'_N} P_{s_1 \dots s_N}$.

Since the interaction in the DK cellular automaton is local, the
tensor $T$ is written as a composite tensor of small local tensors(See
Appendix \ref{app:DK}). Using a diagrammatic notation, we can draw the
two time-steps evolution operator as a network of two types of local
tensors, $B$ and $S$ in \figref{OPTRG_DP}(b). Here, $S$ is a sign
diagonal matrix of which element is 1 or -1.  Since a DK cellular
automaton has a reflection symmetry in a spatial direction, $B$ is
also invariant under reflection.

%
%
\mysection{Tensor renormalization group method with oblique projectors}
The use of a TN representation is expanding not only to equilibrium
systems but also to nonequilibrium ones. For example, a
one-dimensional TN (Matrix Product States) has been used to calculate
a dynamical evolution and a nonequilibrium
steady-state\cite{Johnson:2010gg, Johnson:2015eq, Hotta:2016bc,
  Harada:2019ck}. For the DP, we found that the existence of the
absorbing state governs the unique behavior of informational
entropy\cite{Harada:2019ck}.

Levin and Nave\cite{Levin:2007ju} proposed the first
TRG method, a real-space RG method on a two-dimensional tensor
network, to calculate the partition function of equilibrium
systems. The TRG method assumes an isotropic TN. However,
in general, the TN of nonequilibrium systems is not
isotropic because a time direction is not equal to a spatial
direction.

Xie \textit{et al.}\cite{Xie:2012iy} proposed a simple TRG method
based on a higher-order singular value decomposition(HOSVD) called
HOTRG. Using HOSVD, we determine an optimal orthogonal projector for
each edge of a renormalized tensor. It can be generalized to a
higher-dimensional TN and maybe to an anisotropic one. However, there
is no reflection symmetry in a time direction.  It causes a serious
problem by using an orthogonal projector in HOTRG.  In fact, since the
tensor $B$ in \figref{OPTRG_DP}(b) has no reflection symmetry in a
time direction, an orthogonal projector is not optimal even in the
sense of local optimization. Therefore, we extend it to an oblique
projector that is optimal for a local TN. We can calculate the optimal
oblique projector between two local tensors as in \cite{Corboz:2014ba,
  Iino:2019im}(See Appendix \ref{app:OP}). A simple TRG method with
oblique projectors (OPTRG)\cite{Nataochi:2020} consists of the
coarse-graining of two neighboring tensors by inserting oblique
projectors, as shown in \figref{OPTRG_DP} (c) and (d). We notice that
the coarse-graining tensor in \figref{OPTRG_DP} (c) also keeps a
reflection symmetry in a spatial direction because the original tensor
B has(See Appendix \ref{app:DK}).

In general, nonequilibrium critical systems at critical points are
strongly anisotropic between a spatial direction and a time
direction. The dynamical critical exponent $z$ is not equal to
one. The number of renormalization steps for each direction should not
be the same.  In practice, we should minimize a truncation error in a
renormalization step. Therefore, we always choose a renormalized
direction in which the fidelity between an original tensor and a
renormalized one is larger than in the other direction.

We can calculate a TN representation of a state-probability
distribution at a given time by attaching a TN representation of an
initial distribution to that of transfer probability in
\figref{OPTRG_DP}(b). The expected value of an observable is an inner
product of TN representations of a state-probability distribution and
an observable. We renormalize boundary tensors of an initial
state-probability and an observable with the same projector of a tensor
of transfer probability, $B$. Then, the OPTRG method drastically
improves the accuracy of the TRG calculation of the DK cellular
automaton\cite{Nataochi:2020}.

%
%
\mysection{Renormalized critical tensors of DP}
The renormalized critical tensor in the TRG calculation for
equilibrium critical systems has been well
understood\cite{Gu:2009}. However, we have not studied for
nonequilibrium critical systems yet. We will consider the universal
property of a renormalized critical tensor of the (1+1)-dimensional
DP.

We choose a renormalized direction to reduce a truncation error in the
TRG procedure. \Figref{fig:RG_process} shows the number of TRG steps
in a spatial direction and a time direction of renormalized tensors at
three different critical points of the DK cellular automaton. Here,
the maximum bond dimension $D$ is $80$\footnote{All TRG calculations
  use $D=80$ in this study.}. $n_x$ and $n_t$ denote the number of TRG
steps in a spatial direction and a time direction,
respectively\footnote{We set $n_x=0$, and $n_t=1$ for the initial
  tensor.}. The number of TRG steps increases so that $n_t$ is roughly
proportional to $n_x$. The ratio $n_t/n_x$ at the bond and site DP
critical points is close to the dynamical exponent
$z_{DP}=1.580745(10)$\cite{Jensen:1999jl} of the DP universality
class. The ratio at the compact DP critical point is equal to the
dynamical exponent $z_{CDP}=2$\cite{Dickman:1995ig} of the compact DP
universality class. A renormalized tensor $B$ represents a
$L_t=2^{n_t}$ time-steps evolution on a region of $L_x=2^{n_x}$
sites. If $n_t$ is linearly proportional to $n_x$ roughly with the
coefficient $z$, then $L_x \sim L_t^{1/z}$. As in
\figref{fig:RG_process}, the exponent $z$ is roughly consistent with
the dynamical exponent for the corresponding universality class for
each critical point. The scaling relation corresponds to the scaling
relation of spatial and time correlation lengths at DP critical
points. To reduce the truncation error of renormalization steps, the
aspect ratio of renormalized tensors is automatically proportional to
the ratio of spatial and time correlation lengths at DP critical
points. Therefore, the approximated scaling relation between the
spatial and the time scale of a renormalized tensor is a desired
property of a critical tensor.

\deffig{fig:RG_process}{0.48}{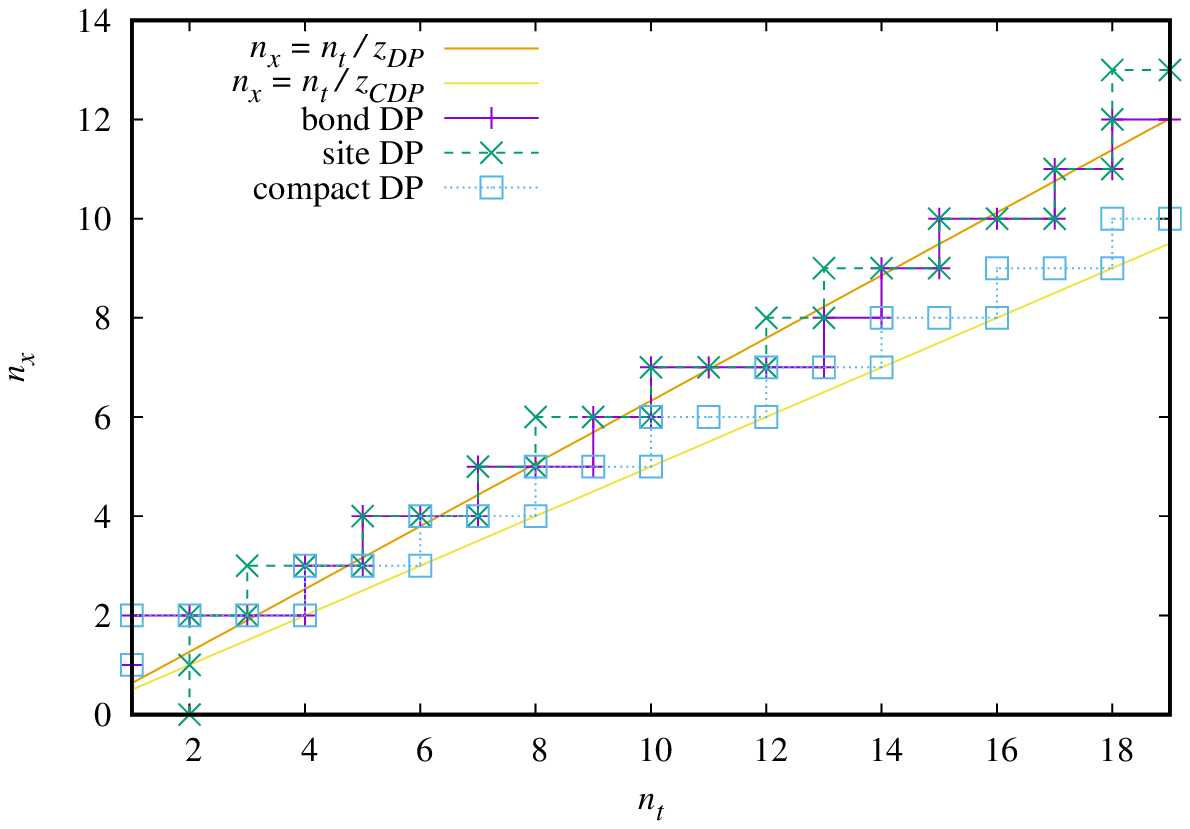}{The number
  of TRG steps in a spatial direction and a time direction at the
  bond, the site, and the compact DP critical points of the DK
  cellular automaton.}

In the case of an equilibrium system, based on a TN representation of
the partition function, we can write the partition function as a trace
of a renormalized tensor and the transfer matrix as a partial trace.
Therefore, a renormalized critical tensor itself has a universal
property of equilibrium critical systems.  In the two-dimensional
system, the universal structure in a spectrum of a renormalized
critical tensor has been well understood, based on a conformal
invariance\cite{Gu:2009}. The rescaled spectrum structure depends on
the universality class, and is related to the scaling dimensions.

\deffig{fig:RG_tensor}{0.25}{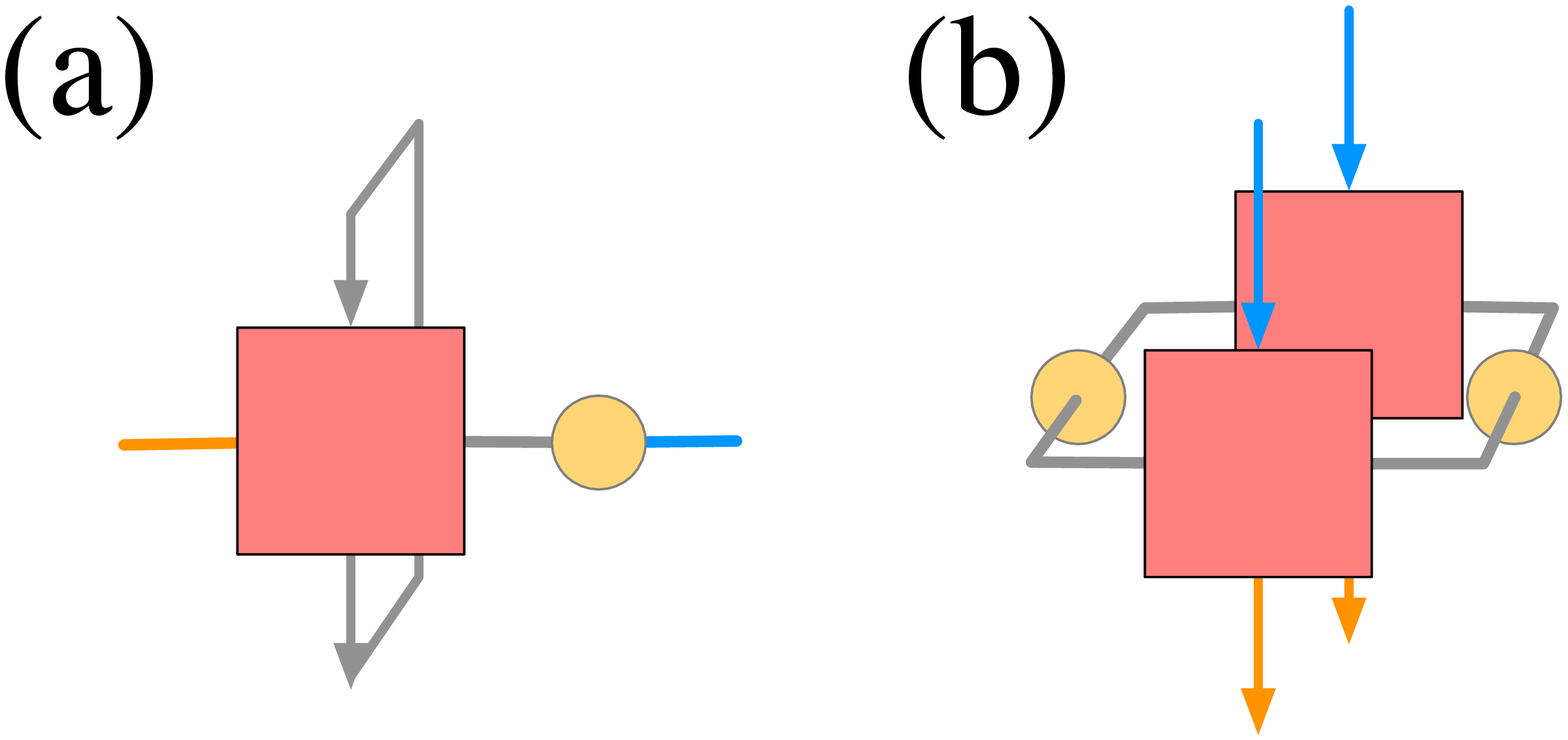}{A transfer
  matrix along (a) a spatial direction and (b) a time direction of a
  renormalized tensor. A (red) box is a renormalized tensor $B$.  An
  (orange) circle is a renormalized sign diagonal matrix.}

Since the DP system is strongly anisotropic, we consider two different
spectra of a renormalized tensor along a spatial direction and a
time direction in \figref{fig:RG_tensor} (a) and (b). If a
renormalized tensor is critical, it is scale-invariant. Then, the
spectrum of a renormalized critical tensor is also
scale-invariant. However, since the aspect ratio of a renormalized
tensor changes in the strongly anisotropic case as DP, the raw
spectrum is not scale-invariant.  Thus, we propose a rescaled spectrum
of a transfer matrix along a spatial direction in
\figref{fig:RG_tensor} (a) as
\begin{equation}
  \label{eq:scaled_spectrum}
  \Delta_{x,i} = -\left(\frac{L_t^{1/z}}{L_x}\right) \log\left|\frac{\lambda_{x,i}}{\lambda_{x,0}}\right|,
\end{equation}
where $\lambda_{x,i}$ is the $i$-th eigenvalue of a transfer matrix
along a spatial direction in descending order and $\lambda_{x,0}$ is
the largest one and $z$ is a dynamical critical exponent. We consider
the absolute value of eigenvalue in \eqref{eq:scaled_spectrum} because
the transfer matrix in \figref{fig:RG_tensor} is generally not
symmetric. Eq. \eqref{eq:scaled_spectrum} is a generalization of a
universal spectrum of a two-dimensional isotropic critical system
proposed in \cite{Gu:2009} with the exponent $1/z$. In the isotropic
case, the rescaled spectrum corresponds to the conformal tower with a
constant factor. \Figref{fig:es_x} shows the rescaled spectrum of
renormalized tensors for three different critical DPs. The rescaled
spectrum is scale-invariant in the wide range from $L_t = 2^5$ to
$2^{14}$ for all critical points. Since the accuracy of a higher
spectrum becomes a bit unstable after a TRG procedure along a spatial
direction, we only plot the results after the TRG along a time
direction. The correlation length along a spatial direction,
$\xi_{\perp}$, is proportional to $L_t^{1/z}$ after the $L_t$
time-steps evolution. Thus, the scale-invariance of $\Delta_{x,i}$ in
\eqref{eq:scaled_spectrum} is consistent with the expected form of
eigenvalues of a transfer matrix along a spatial direction as
$|\lambda_{x,i}| \propto \exp[-c_i L_x/\xi_{\perp}]$.

\deffig{fig:es_x}{0.48}{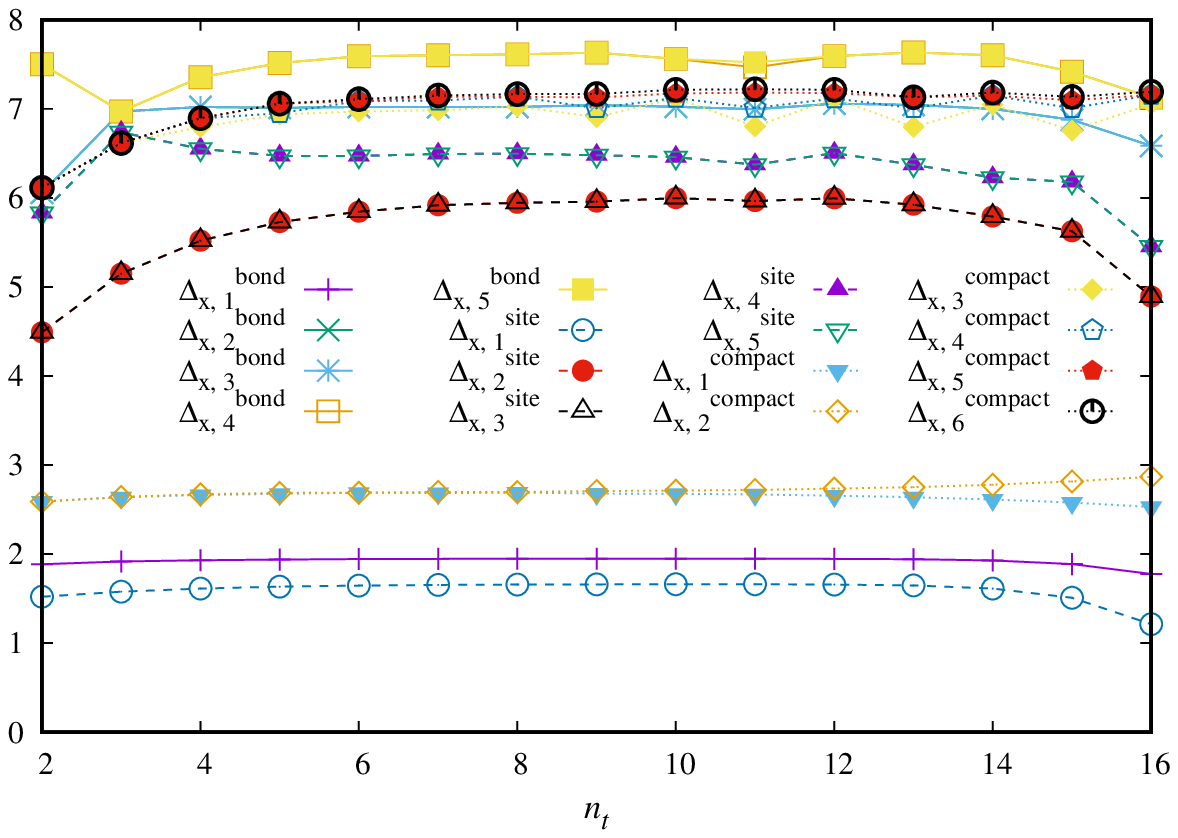}{Rescaled spectrum of a
  transfer matrix along a spatial direction in
  \figref{fig:RG_tensor}(a).}

Some spectra are degenerate in \figref{fig:es_x}. The degeneracy of
the spectrum in the bond DP is equal to that in the site DP as
$1, 2, 4, \cdots$ in the ascending level from $\Delta_{x,1}$. On the
other hand, the degeneracy of the lowest level in the compact DP is 2,
not 1.  Therefore, the degeneracy depends on the universality
class. The degeneracy of the second level in the compact DP is unclear
and the degeneracy of the lowest level is slightly violated for
renormalized critical tensors larger than $L_t \ge 2^{11}$ because the
accuracy of renormalized critical tensors is low. The eigenvalues
$\lambda_{x,0}$ and $\lambda_{x,1}$ of the bond and the site DPs are
real, but the others are complex in \figref{fig:es_x}. The eigenvalues
$\Delta_{x,0}$, $\Delta_{x,1}$, and $\Delta_{x,2}$ of the compact DP
are real, but the others are complex. The level interval of the bond
DP is proportional to that of the site DP. We summarize the values of
rescaled spectra and the ratios in Tab. \ref{tab:es_x}, which is based
on the data from $n_t=8$ to $11$ for the bond and the site DPs and at
$n_t=8, 10$, and $12$ for the compact DP. The ratios of the second level
to the first level for the bond and the site DPs,
$\Delta_{x, 2}/\Delta_{x,1}$, are 3.60(1) and 3.59(1), respectively.
They agree within error bars. But they do not agree with the compact
DP's ratio. The ratios of the fourth level to the first level for the
bond and the site DPs, $\Delta_{x, 4}/\Delta_{x,1}$, also agree within
error bars. Therefore, the rescaled spectrum structure is universal.

\begin{table}[h]
  \centering
  \caption{Values and ratios of rescaled spectra along a spatial
    direction.}
  \begin{tabular}{|c|c|c|c|}
    \hline
    & $\Delta_{x,1}$ & $\Delta_{x, 2\sim 3}/\Delta_{x,1}$ & $\Delta_{x, 4\sim 7}/\Delta_{x,1}$\\
    \hline
    bond DP & 1.948(1) & 3.60(1) & 3.91(2)\\
    site DP & 1.660(2) & 3.59(1) & 3.91(1)\\
    \hline
    & $\Delta_{x,1\sim 2}$ & $\Delta_{x,3} / \Delta_{x,1\sim 2}$  & \multicolumn{1}{c}{}\\
    \cline{1-3}
    compact DP & 2.69(2) & 2.65(4)  & \multicolumn{1}{c}{}\\
    \cline{1-3}
  \end{tabular}
  \label{tab:es_x}
\end{table}

For the time direction, we propose a rescaled spectrum of a transfer
matrix in \figref{fig:RG_tensor} (b) as
\begin{equation}
  \label{eq:es_t}
  \Delta_{t,i} = -\left(\frac{L_x^z}{L_t}\right) \log\left|\frac{\lambda_{t,i}}{\lambda_{t,0}}\right|,
\end{equation}
where $\lambda_{t,i}$ is the $i$-th eigenvalue of a transfer matrix
along a time direction in descending order and $\lambda_{t,0}$ is the
largest one and $z$ is a dynamical critical exponent. As in
\Figref{fig:es_t}, $\Delta_{t,i}$ is scale-invariant in the wide range
for all critical points. All eigenvalues in $\Delta_{t,i}$ of
\figref{fig:es_t} are real. The scale-invariance of $\Delta_{t,i}$ in
\eqref{eq:es_t} is consistent with the expected form of eigenvalues of
a transfer matrix along a time direction because
$\xi_{\parallel} \propto L_x^z$. We summarize the values of rescaled
spectra and the ratios in Tab. \ref{tab:es_t}, which is based on the
data from $n_t=8$ to $10$ for the bond and the site DPs and at
$n_t=8, 10$, and $12$ for the compact DP.  $\Delta_{t, 3}$ and
$\Delta_{t, 4}$ are degenerate in the bond and the site DPs. The first
spectrum $\Delta_{t,1}$ of the bond and the site DPs are small, and
their accuracy are low. Thus, we compare the ratios
$\Delta_{t, 3\sim 4}/\Delta_{t, 2}$ for both DPs. As in
Tab. \ref{tab:es_t}, they agree within error bars. However, the
spectrum structure of the compact DP is different from that of the
bond and the site DPs. The first eigenvalue $\lambda_{t, 1}$ is
degenerate to the largest eigenvalue, $\lambda_{t, 0}$. It is
consistent with the existence of two absorbing states in the compact
DP. The other degeneracy is also different from that of the bond and
the site DPs. Interestingly, the ratio of each level to the second
level of the compact DP is an integer within error bars. As in
Tab. \ref{tab:es_t}, the rescaled spectrum structures depend on the
universality class. Therefore, the rescaled spectrum structure of a
transfer matrix along a time direction is universal.

\deffig{fig:es_t}{0.48}{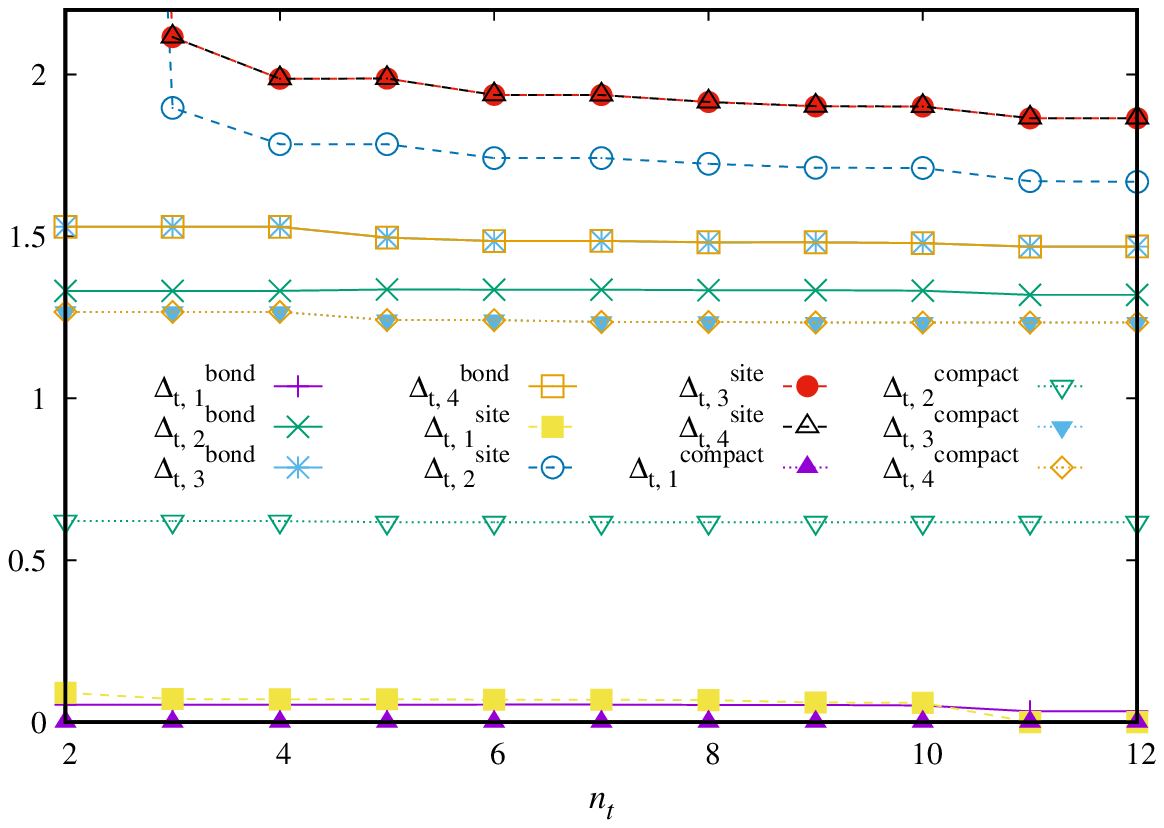}{Rescaled spectrum of a
  transfer matrix along a time direction in \figref{fig:RG_tensor}
  (b).}

\begin{table}[h]
  \centering
  \caption{Values and ratios of rescaled spectra along a time
    direction.}
  \begin{tabular}{|c|c|c|c|c|}
    \cline{1-4}
    & $\Delta_{t,1}$ & $\Delta_{t,2}$ & $\Delta_{t, 3\sim 4}/\Delta_{t,2}$& \multicolumn{1}{c}{}\\
    \cline{1-4}
    bond DP & 0.053(2) & 1.334(2) & 1.110(1) & \multicolumn{1}{c}{}\\
    site DP & 0.063(5) & 1.716(8) & 1.111(1) & \multicolumn{1}{c}{}\\
    \hline
    & $\Delta_{t, 2}$ & $\Delta_{t, 3\sim 4}/\Delta_{t, 2}$ & $\Delta_{t, 5}/\Delta_{t, 2}$ & $\Delta_{t, 6 \sim 9}/\Delta_{t, 2}$ \\
    \hline
    compact DP & 0.6172(4) & 2.000(2) & 4.000(4) & 5.004(5)\\
    \hline
  \end{tabular}
  \label{tab:es_t}
\end{table}

%
%
\mysection{Conclusion}
To understand the universality of nonequilibrium critical systems, we
tried the numerical RG approach for the one-dimensional DK stochastic
cellular automaton that is a dynamical model of the (1+1)-dimensional
DP. Using a TRG method with oblique projectors for the TN
representation of the time evolution operator, we numerically
calculate renormalized critical tensors for the bond, the site, and
the compact DPs. We proposed the rescaled spectrum of a renormalized
critical tensor with a strongly anisotropic criticality. We
numerically confirmed that the rescaled spectra of renormalized
critical tensors at nonequilibrium critical points of the
(1+1)-dimensional DP are scale-invariant and the rescaled spectrum
structure is universal. Future works need to understand the universal
spectrum structure. We calculated the RG fixed points in the tensor
space for the (1+1)-dimensional DP. Our approach is applicable to the
other nonequilibrium criticalities or the higher dimensional cases. To
improve the accuracy of renormalization steps, the generalization of
techniques in \cite{Evenbly:2015cs, Yang:2017hj, Hauru:2018ij,
  Harada:2018cj} for the strongly anisotropic criticality is
promising.

\begin{acknowledgments}
K.H appreciates fruitful comments from N. Kawashima. This work was
supported by JSPS KAKENHI Grant No. 17K05576 and No. 20K03766. The
computation in this work has been done using the facilities of the
Supercomputer Center, the Institute for Solid State Physics, the
University of Tokyo and the facilities of the Supercomputer Center at
Kyoto University.
\end{acknowledgments}

\newpage
\appendix
\section{Tensor network representation of the time evolution operator
  of the one-dimensional Domany-Kinzel cellular automaton}
\label{app:DK}
Since the interaction in the one-dimensional Domany-Kinzel(DK)
cellular automaton\cite{Domany:1984fh} consists is local, the transfer
probability tensor $T$ is written as a composite tensor of small local
tensors. Using a diagrammatic notation, we can draw it as a network of
two types of local tensors, $w$ and $\delta$, in \figref{TN_DP}(a),
which is a tensor network(TN) representation of $T$. Here, a local
tensor is a graphical object with edges. An edge denotes the index of
a local tensor, and a connected edge means a tensor contraction
between the corresponding indexes of two local tensors. The definition
of local tensors, $w$ and $\delta$, is as follows:
\begin{align*}
  w^{lm}_n &= (1-n) + (2n-1)P[l+m].\\
  \delta^n_{op} & =
            \begin{cases}
              1 & \mbox{if $n=o=p$.}\\
              0 & \mbox{otherwise.}
            \end{cases}
\end{align*}
Here, $P[0], P[1]$, and $P[2]$ are free parameters which define the DK
cellular automaton. We introduce a new local tensor $A$ as
$A^{lm}_{op} = \sum_n \delta^n_{op}w^{lm}_n$ in
\figref{TN_DP}(a). Since it can be transformed into a symmetric matrix
$A_{(lo), (mp)}$ along a horizontal direction, it can be diagonalized
as $A=VSV^t$. Here, $S$ is a sign diagonal matrix of which diagonal
element is 1 or -1. Combining four local tensors $\delta, w, V$, and
$V^t$, we introduce a new tensor $B$ in \figref{TN_DP}(b). Then, the
two time-steps evolution operator can be transformed into the TN of
$B$ and $S$.
\deffig{TN_DP}{0.48}{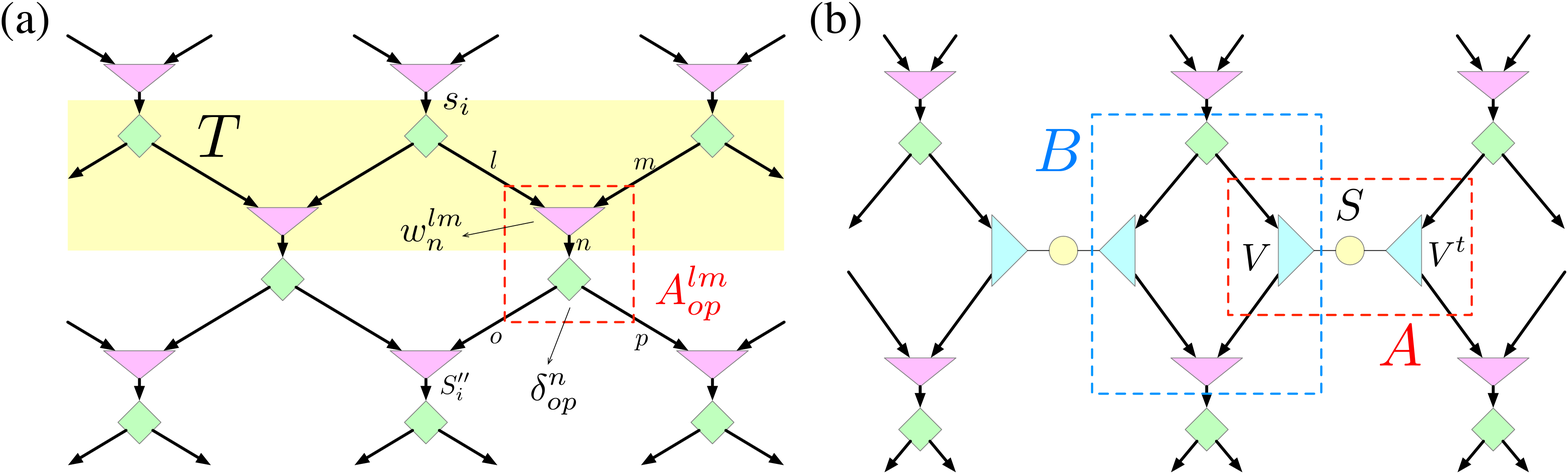}{(a) TN representation of a transfer
  probability tensor $T$ in the (yellow) shaded box. The new local
  tensor $A$ has a reflection symmetry along a horizontal direction
  (b) Decomposition of $A$ and a new local tensor $B$ and $S$. }
\section{Optimal oblique projector}
\label{app:OP}
We consider an approximation of a matrix product of \figref{OP}(a)
into \figref{OP}(b) with a reduced bond dimension. To minimize the
Frobenius norm between them under a fixed bond dimension $\chi'$, we
keep the $\chi'$ largest singular values of $ASB$ by inserting an
oblique projector $PS'Q$ between $A$ and $B$. Such $P, Q$ and $S'$ can
be calculated as the following procedures\cite{Corboz:2014ba,
  Iino:2019im}.
\begin{enumerate}
\item Singular value decomposition: $M=ASB=U \Lambda V^\dagger$, where
  $U$ and $V$ are unitary, $\Lambda$ is a diagonal matrix of singular
  values.
\item Extract the $\chi'$ largest components of singular values:
  $\Lambda \to \tilde{\Lambda}$, $U \to \tilde{U}$, $V \to \tilde{V}$.
\item
  $P=S B \tilde{V} \vert \tilde{\Lambda} \vert^{-\frac12},\ 
  Q = \vert \tilde{\Lambda} \vert^{-\frac12} \tilde{U}^\dagger A S,\ 
  S' = \mbox{sign}(\tilde{\Lambda})$.
\end{enumerate}
If a matrix $ASB$ is Hermitian, $B = A^\dagger$, then
$M=U\Lambda U^\dagger$ and $P=Q^\dagger$.

\deffig{OP}{0.48}{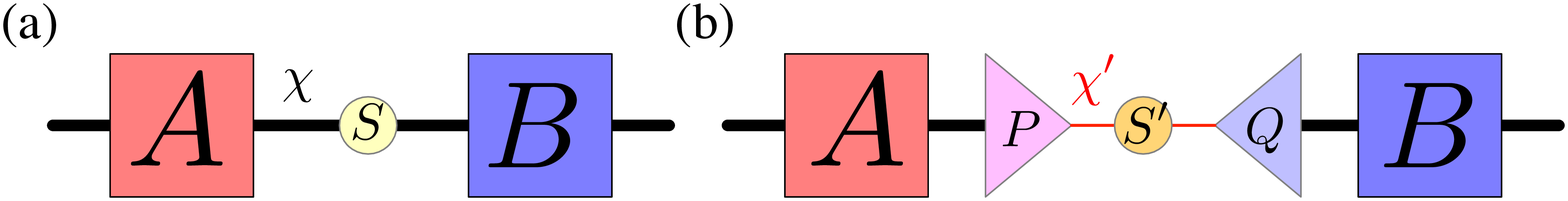}{(a) Matrix product with a sign diagonal
  matrix. (b) Reducing a bond dimension from $\chi$ to $\chi'$ with an
  oblique projector with a new sign diagonal matrix.}
\newpage
\bibliography{dp}

\begin{thebibliography}{27}%
\makeatletter
\providecommand \@ifxundefined [1]{%
 \@ifx{#1\undefined}
}%
\providecommand \@ifnum [1]{%
 \ifnum #1\expandafter \@firstoftwo
 \else \expandafter \@secondoftwo
 \fi
}%
\providecommand \@ifx [1]{%
 \ifx #1\expandafter \@firstoftwo
 \else \expandafter \@secondoftwo
 \fi
}%
\providecommand \natexlab [1]{#1}%
\providecommand \enquote  [1]{``#1''}%
\providecommand \bibnamefont  [1]{#1}%
\providecommand \bibfnamefont [1]{#1}%
\providecommand \citenamefont [1]{#1}%
\providecommand \href@noop [0]{\@secondoftwo}%
\providecommand \href [0]{\begingroup \@sanitize@url \@href}%
\providecommand \@href[1]{\@@startlink{#1}\@@href}%
\providecommand \@@href[1]{\endgroup#1\@@endlink}%
\providecommand \@sanitize@url [0]{\catcode `\\12\catcode `\$12\catcode
  `\&12\catcode `\#12\catcode `\^12\catcode `\_12\catcode `\%12\relax}%
\providecommand \@@startlink[1]{}%
\providecommand \@@endlink[0]{}%
\providecommand \url  [0]{\begingroup\@sanitize@url \@url }%
\providecommand \@url [1]{\endgroup\@href {#1}{\urlprefix }}%
\providecommand \urlprefix  [0]{URL }%
\providecommand \Eprint [0]{\href }%
\providecommand \doibase [0]{http://dx.doi.org/}%
\providecommand \selectlanguage [0]{\@gobble}%
\providecommand \bibinfo  [0]{\@secondoftwo}%
\providecommand \bibfield  [0]{\@secondoftwo}%
\providecommand \translation [1]{[#1]}%
\providecommand \BibitemOpen [0]{}%
\providecommand \bibitemStop [0]{}%
\providecommand \bibitemNoStop [0]{.\EOS\space}%
\providecommand \EOS [0]{\spacefactor3000\relax}%
\providecommand \BibitemShut  [1]{\csname bibitem#1\endcsname}%
\let\auto@bib@innerbib\@empty
\bibitem [{\citenamefont {Takeuchi}\ \emph {et~al.}(2007)\citenamefont
  {Takeuchi}, \citenamefont {Kuroda}, \citenamefont {Chat{\'e}},\ and\
  \citenamefont {Sano}}]{Takeuchi:2007co}%
  \BibitemOpen
  \bibfield  {author} {\bibinfo {author} {\bibfnamefont {K.A.}\ \bibnamefont
  {Takeuchi}}, \bibinfo {author} {\bibfnamefont {M.}~\bibnamefont {Kuroda}},
  \bibinfo {author} {\bibfnamefont {H.}~\bibnamefont {Chat{\'e}}}, \ and\
  \bibinfo {author} {\bibfnamefont {M.}~\bibnamefont {Sano}},\ }\bibfield
  {title} {\enquote {\bibinfo {title} {{Directed Percolation Criticality in
  Turbulent Liquid Crystals}},}\ }\href@noop {} {\bibfield  {journal} {\bibinfo
   {journal} {Physical Review Letters}\ }\textbf {\bibinfo {volume} {99}},\
  \bibinfo {pages} {234503} (\bibinfo {year} {2007})}\BibitemShut {NoStop}%
\bibitem [{\citenamefont {Takeuchi}\ \emph {et~al.}(2009)\citenamefont
  {Takeuchi}, \citenamefont {Kuroda}, \citenamefont {Chat{\'e}},\ and\
  \citenamefont {Sano}}]{Takeuchi:2009ia}%
  \BibitemOpen
  \bibfield  {author} {\bibinfo {author} {\bibfnamefont {K.A.}\ \bibnamefont
  {Takeuchi}}, \bibinfo {author} {\bibfnamefont {M.}~\bibnamefont {Kuroda}},
  \bibinfo {author} {\bibfnamefont {H.}~\bibnamefont {Chat{\'e}}}, \ and\
  \bibinfo {author} {\bibfnamefont {M.}~\bibnamefont {Sano}},\ }\bibfield
  {title} {\enquote {\bibinfo {title} {{Experimental realization of directed
  percolation criticality in turbulent liquid crystals}},}\ }\href@noop {}
  {\bibfield  {journal} {\bibinfo  {journal} {Physical Review E}\ }\textbf
  {\bibinfo {volume} {80}},\ \bibinfo {pages} {051116} (\bibinfo {year}
  {2009})}\BibitemShut {NoStop}%
\bibitem [{\citenamefont {Sano}\ and\ \citenamefont
  {Tamai}(2016)}]{Sano:2016kh}%
  \BibitemOpen
  \bibfield  {author} {\bibinfo {author} {\bibfnamefont {Masaki}\ \bibnamefont
  {Sano}}\ and\ \bibinfo {author} {\bibfnamefont {Keiichi}\ \bibnamefont
  {Tamai}},\ }\bibfield  {title} {\enquote {\bibinfo {title} {{A universal
  transition to turbulence in channel~flow}},}\ }\href@noop {} {\bibfield
  {journal} {\bibinfo  {journal} {Nature Physics}\ }\textbf {\bibinfo {volume}
  {12}},\ \bibinfo {pages} {249--253} (\bibinfo {year} {2016})}\BibitemShut
  {NoStop}%
\bibitem [{\citenamefont {Henkel}\ \emph {et~al.}(2008)\citenamefont {Henkel},
  \citenamefont {Hinrichsen},\ and\ \citenamefont
  {L{\"u}beck}}]{Henkel:2008vv}%
  \BibitemOpen
  \bibfield  {author} {\bibinfo {author} {\bibfnamefont {M.}~\bibnamefont
  {Henkel}}, \bibinfo {author} {\bibfnamefont {H.}~\bibnamefont {Hinrichsen}},
  \ and\ \bibinfo {author} {\bibfnamefont {S.}~\bibnamefont {L{\"u}beck}},\
  }\href@noop {} {\emph {\bibinfo {title} {{Non-Equilibrium Phase
  Transitions}}}},\ Vol.~\bibinfo {volume} {1}\ (\bibinfo  {publisher}
  {Springer},\ \bibinfo {address} {New York},\ \bibinfo {year}
  {2008})\BibitemShut {NoStop}%
\bibitem [{\citenamefont {Wilson}(1971{\natexlab{a}})}]{Wilson:1971fb}%
  \BibitemOpen
  \bibfield  {author} {\bibinfo {author} {\bibfnamefont {Kenneth~G.}\
  \bibnamefont {Wilson}},\ }\bibfield  {title} {\enquote {\bibinfo {title}
  {{Renormalization Group and Critical Phenomena. I. Renormalization Group and
  the Kadanoff Scaling Picture}},}\ }\href {\doibase 10.1103/PhysRevB.4.3174}
  {\bibfield  {journal} {\bibinfo  {journal} {Physical Review B}\ }\textbf
  {\bibinfo {volume} {4}},\ \bibinfo {pages} {3174--3183} (\bibinfo {year}
  {1971}{\natexlab{a}})}\BibitemShut {NoStop}%
\bibitem [{\citenamefont {Wilson}(1971{\natexlab{b}})}]{Wilson:1971bg}%
  \BibitemOpen
  \bibfield  {author} {\bibinfo {author} {\bibfnamefont {Kenneth~G}\
  \bibnamefont {Wilson}},\ }\bibfield  {title} {\enquote {\bibinfo {title}
  {{Renormalization Group and Critical Phenomena. II. Phase-Space Cell Analysis
  of Critical Behavior}},}\ }\href {\doibase 10.1103/PhysRevB.4.3184}
  {\bibfield  {journal} {\bibinfo  {journal} {Physical Review B}\ }\textbf
  {\bibinfo {volume} {4}},\ \bibinfo {pages} {3184--3205} (\bibinfo {year}
  {1971}{\natexlab{b}})}\BibitemShut {NoStop}%
\bibitem [{\citenamefont {Cardy}(1996)}]{cardy_1996}%
  \BibitemOpen
  \bibfield  {author} {\bibinfo {author} {\bibfnamefont {John}\ \bibnamefont
  {Cardy}},\ }\href@noop {} {\emph {\bibinfo {title} {Scaling and
  Renormalization in Statistical Physics}}},\ Cambridge Lecture Notes in
  Physics\ (\bibinfo  {publisher} {Cambridge University Press},\ \bibinfo
  {year} {1996})\BibitemShut {NoStop}%
\bibitem [{\citenamefont {Goldenfeld}(2018)}]{goldenfeld2018lectures}%
  \BibitemOpen
  \bibfield  {author} {\bibinfo {author} {\bibfnamefont {Nigel}\ \bibnamefont
  {Goldenfeld}},\ }\href@noop {} {\emph {\bibinfo {title} {Lectures On Phase
  Transitions And The Renormalization Group}}}\ (\bibinfo  {publisher} {CRC
  Press},\ \bibinfo {year} {2018})\BibitemShut {NoStop}%
\bibitem [{\citenamefont {Gu}\ and\ \citenamefont {Wen}(2009)}]{Gu:2009}%
  \BibitemOpen
  \bibfield  {author} {\bibinfo {author} {\bibfnamefont {Zheng-Cheng}\
  \bibnamefont {Gu}}\ and\ \bibinfo {author} {\bibfnamefont {Xiao-Gang}\
  \bibnamefont {Wen}},\ }\bibfield  {title} {\enquote {\bibinfo {title}
  {{Tensor-entanglement-filtering renormalization approach and
  symmetry-protected topological order}},}\ }\href {\doibase
  10.1103/PhysRevB.80.155131} {\bibfield  {journal} {\bibinfo  {journal}
  {Physical Review B}\ }\textbf {\bibinfo {volume} {80}},\ \bibinfo {pages}
  {155131} (\bibinfo {year} {2009})}\BibitemShut {NoStop}%
\bibitem [{\citenamefont {Domany}\ and\ \citenamefont
  {Kinzel}(1984)}]{Domany:1984fh}%
  \BibitemOpen
  \bibfield  {author} {\bibinfo {author} {\bibfnamefont {Eythan}\ \bibnamefont
  {Domany}}\ and\ \bibinfo {author} {\bibfnamefont {Wolfgang}\ \bibnamefont
  {Kinzel}},\ }\bibfield  {title} {\enquote {\bibinfo {title} {{Equivalence of
  Cellular Automata to Ising Models and Directed Percolation}},}\ }\href@noop
  {} {\bibfield  {journal} {\bibinfo  {journal} {Physical Review Letters}\
  }\textbf {\bibinfo {volume} {53}},\ \bibinfo {pages} {311--314} (\bibinfo
  {year} {1984})}\BibitemShut {NoStop}%
\bibitem [{\citenamefont {Johnson}\ \emph {et~al.}(2010)\citenamefont
  {Johnson}, \citenamefont {Clark},\ and\ \citenamefont
  {Jaksch}}]{Johnson:2010gg}%
  \BibitemOpen
  \bibfield  {author} {\bibinfo {author} {\bibfnamefont {T.H.}\ \bibnamefont
  {Johnson}}, \bibinfo {author} {\bibfnamefont {S.R.}\ \bibnamefont {Clark}}, \
  and\ \bibinfo {author} {\bibfnamefont {D.}~\bibnamefont {Jaksch}},\
  }\bibfield  {title} {\enquote {\bibinfo {title} {{Dynamical simulations of
  classical stochastic systems using matrix product states}},}\ }\href@noop {}
  {\bibfield  {journal} {\bibinfo  {journal} {Physical Review E}\ }\textbf
  {\bibinfo {volume} {82}},\ \bibinfo {pages} {036702} (\bibinfo {year}
  {2010})}\BibitemShut {NoStop}%
\bibitem [{\citenamefont {Johnson}\ \emph {et~al.}(2015)\citenamefont
  {Johnson}, \citenamefont {Elliott}, \citenamefont {Clark},\ and\
  \citenamefont {Jaksch}}]{Johnson:2015eq}%
  \BibitemOpen
  \bibfield  {author} {\bibinfo {author} {\bibfnamefont {T.H.}\ \bibnamefont
  {Johnson}}, \bibinfo {author} {\bibfnamefont {T.J.}\ \bibnamefont {Elliott}},
  \bibinfo {author} {\bibfnamefont {S.R.}\ \bibnamefont {Clark}}, \ and\
  \bibinfo {author} {\bibfnamefont {D.}~\bibnamefont {Jaksch}},\ }\bibfield
  {title} {\enquote {\bibinfo {title} {{Capturing Exponential Variance Using
  Polynomial Resources: Applying Tensor Networks to Nonequilibrium Stochastic
  Processes}},}\ }\href@noop {} {\bibfield  {journal} {\bibinfo  {journal}
  {Physical Review Letters}\ }\textbf {\bibinfo {volume} {114}},\ \bibinfo
  {pages} {090602} (\bibinfo {year} {2015})}\BibitemShut {NoStop}%
\bibitem [{\citenamefont {Hotta}(2016)}]{Hotta:2016bc}%
  \BibitemOpen
  \bibfield  {author} {\bibinfo {author} {\bibfnamefont {Yoshihiko}\
  \bibnamefont {Hotta}},\ }\bibfield  {title} {\enquote {\bibinfo {title}
  {{Tensor-network algorithm for nonequilibrium relaxation in the thermodynamic
  limit}},}\ }\href@noop {} {\bibfield  {journal} {\bibinfo  {journal}
  {Physical Review E}\ }\textbf {\bibinfo {volume} {93}},\ \bibinfo {pages}
  {062136} (\bibinfo {year} {2016})}\BibitemShut {NoStop}%
\bibitem [{\citenamefont {Harada}\ and\ \citenamefont
  {Kawashima}(2019)}]{Harada:2019ck}%
  \BibitemOpen
  \bibfield  {author} {\bibinfo {author} {\bibfnamefont {Kenji}\ \bibnamefont
  {Harada}}\ and\ \bibinfo {author} {\bibfnamefont {Naoki}\ \bibnamefont
  {Kawashima}},\ }\bibfield  {title} {\enquote {\bibinfo {title} {{Entropy
  Governed by the Absorbing State of Directed Percolation}},}\ }\href {\doibase
  10.1103/PhysRevLett.123.090601} {\bibfield  {journal} {\bibinfo  {journal}
  {Physical Review Letters}\ }\textbf {\bibinfo {volume} {123}},\ \bibinfo
  {pages} {090601} (\bibinfo {year} {2019})}\BibitemShut {NoStop}%
\bibitem [{\citenamefont {Levin}\ and\ \citenamefont
  {Nave}(2007)}]{Levin:2007ju}%
  \BibitemOpen
  \bibfield  {author} {\bibinfo {author} {\bibfnamefont {Michael}\ \bibnamefont
  {Levin}}\ and\ \bibinfo {author} {\bibfnamefont {Cody~P.}\ \bibnamefont
  {Nave}},\ }\bibfield  {title} {\enquote {\bibinfo {title} {{Tensor
  Renormalization Group Approach to Two-Dimensional Classical Lattice
  Models}},}\ }\href {\doibase 10.1103/PhysRevLett.99.120601} {\bibfield
  {journal} {\bibinfo  {journal} {Physical Review Letters}\ }\textbf {\bibinfo
  {volume} {99}},\ \bibinfo {pages} {120601} (\bibinfo {year}
  {2007})}\BibitemShut {NoStop}%
\bibitem [{\citenamefont {Xie}\ \emph {et~al.}(2012)\citenamefont {Xie},
  \citenamefont {Chen}, \citenamefont {Qin}, \citenamefont {Zhu}, \citenamefont
  {Yang},\ and\ \citenamefont {Xiang}}]{Xie:2012iy}%
  \BibitemOpen
  \bibfield  {author} {\bibinfo {author} {\bibfnamefont {Z.~Y.}\ \bibnamefont
  {Xie}}, \bibinfo {author} {\bibfnamefont {J.}~\bibnamefont {Chen}}, \bibinfo
  {author} {\bibfnamefont {M.~P.}\ \bibnamefont {Qin}}, \bibinfo {author}
  {\bibfnamefont {J.~W.}\ \bibnamefont {Zhu}}, \bibinfo {author} {\bibfnamefont
  {L.~P.}\ \bibnamefont {Yang}}, \ and\ \bibinfo {author} {\bibfnamefont
  {T.}~\bibnamefont {Xiang}},\ }\bibfield  {title} {\enquote {\bibinfo {title}
  {{Coarse-graining renormalization by higher-order singular value
  decomposition}},}\ }\href {\doibase 10.1103/PhysRevB.86.045139} {\bibfield
  {journal} {\bibinfo  {journal} {Physical Review B}\ }\textbf {\bibinfo
  {volume} {86}},\ \bibinfo {pages} {045139} (\bibinfo {year}
  {2012})}\BibitemShut {NoStop}%
\bibitem [{\citenamefont {Corboz}\ \emph {et~al.}(2014)\citenamefont {Corboz},
  \citenamefont {Rice},\ and\ \citenamefont {Troyer}}]{Corboz:2014ba}%
  \BibitemOpen
  \bibfield  {author} {\bibinfo {author} {\bibfnamefont {Philippe}\
  \bibnamefont {Corboz}}, \bibinfo {author} {\bibfnamefont {T.~M.}\
  \bibnamefont {Rice}}, \ and\ \bibinfo {author} {\bibfnamefont {Matthias}\
  \bibnamefont {Troyer}},\ }\bibfield  {title} {\enquote {\bibinfo {title}
  {{Competing States in the t-J Model: Uniform d-Wave State versus Stripe
  State}},}\ }\href {\doibase 10.1103/PhysRevLett.113.046402} {\bibfield
  {journal} {\bibinfo  {journal} {Physical Review Letters}\ }\textbf {\bibinfo
  {volume} {113}},\ \bibinfo {pages} {046402} (\bibinfo {year}
  {2014})}\BibitemShut {NoStop}%
\bibitem [{\citenamefont {Iino}\ \emph {et~al.}(2019)\citenamefont {Iino},
  \citenamefont {Morita},\ and\ \citenamefont {Kawashima}}]{Iino:2019im}%
  \BibitemOpen
  \bibfield  {author} {\bibinfo {author} {\bibfnamefont {Shumpei}\ \bibnamefont
  {Iino}}, \bibinfo {author} {\bibfnamefont {Satoshi}\ \bibnamefont {Morita}},
  \ and\ \bibinfo {author} {\bibfnamefont {Naoki}\ \bibnamefont {Kawashima}},\
  }\bibfield  {title} {\enquote {\bibinfo {title} {{Boundary tensor
  renormalization group}},}\ }\href {\doibase 10.1103/PhysRevB.100.035449}
  {\bibfield  {journal} {\bibinfo  {journal} {Physical Review B}\ }\textbf
  {\bibinfo {volume} {100}},\ \bibinfo {pages} {035449} (\bibinfo {year}
  {2019})}\BibitemShut {NoStop}%
\bibitem [{\citenamefont {Nataochi}\ and\ \citenamefont
  {Harada}()}]{Nataochi:2020}%
  \BibitemOpen
  \bibfield  {author} {\bibinfo {author} {\bibfnamefont {Kana}\ \bibnamefont
  {Nataochi}}\ and\ \bibinfo {author} {\bibfnamefont {Kenji}\ \bibnamefont
  {Harada}},\ }\href@noop {} {\enquote {\bibinfo {title} {{Tensor
  renormalization group method with oblique projectors for a dynamical
  process}},}\ }\bibinfo {note} {{in preparation.}}\BibitemShut {Stop}%
\bibitem [{Note1()}]{Note1}%
  \BibitemOpen
  \bibinfo {note} {All TRG calculations use $D=80$ in this study.}\BibitemShut
  {Stop}%
\bibitem [{Note2()}]{Note2}%
  \BibitemOpen
  \bibinfo {note} {We set $n_x=0$, and $n_t=1$ for the initial
  tensor.}\BibitemShut {Stop}%
\bibitem [{\citenamefont {Jensen}(1999)}]{Jensen:1999jl}%
  \BibitemOpen
  \bibfield  {author} {\bibinfo {author} {\bibfnamefont {Iwan}\ \bibnamefont
  {Jensen}},\ }\bibfield  {title} {\enquote {\bibinfo {title} {{Low-density
  series expansions for directed percolation: I. A new efficient algorithm with
  applications to the square lattice}},}\ }\href@noop {} {\bibfield  {journal}
  {\bibinfo  {journal} {Journal of Physics A: Mathematical and General}\
  }\textbf {\bibinfo {volume} {32}},\ \bibinfo {pages} {5233--5249} (\bibinfo
  {year} {1999})}\BibitemShut {NoStop}%
\bibitem [{\citenamefont {Dickman}\ and\ \citenamefont
  {Tretyakov}(1995)}]{Dickman:1995ig}%
  \BibitemOpen
  \bibfield  {author} {\bibinfo {author} {\bibfnamefont {Ronald}\ \bibnamefont
  {Dickman}}\ and\ \bibinfo {author} {\bibfnamefont {Alex~Yu.}\ \bibnamefont
  {Tretyakov}},\ }\bibfield  {title} {\enquote {\bibinfo {title} {{Hyperscaling
  in the Domany-Kinzel cellular automaton}},}\ }\href {\doibase
  10.1103/PhysRevE.52.3218} {\bibfield  {journal} {\bibinfo  {journal}
  {Physical Review E}\ }\textbf {\bibinfo {volume} {52}},\ \bibinfo {pages}
  {3218--3220} (\bibinfo {year} {1995})}\BibitemShut {NoStop}%
\bibitem [{\citenamefont {Evenbly}\ and\ \citenamefont
  {Vidal}(2015)}]{Evenbly:2015cs}%
  \BibitemOpen
  \bibfield  {author} {\bibinfo {author} {\bibfnamefont {G.}~\bibnamefont
  {Evenbly}}\ and\ \bibinfo {author} {\bibfnamefont {G.}~\bibnamefont
  {Vidal}},\ }\bibfield  {title} {\enquote {\bibinfo {title} {{Tensor Network
  Renormalization}},}\ }\href {\doibase 10.1103/PhysRevLett.115.180405}
  {\bibfield  {journal} {\bibinfo  {journal} {Physical Review Letters}\
  }\textbf {\bibinfo {volume} {115}},\ \bibinfo {pages} {180405} (\bibinfo
  {year} {2015})}\BibitemShut {NoStop}%
\bibitem [{\citenamefont {Yang}\ \emph {et~al.}(2017)\citenamefont {Yang},
  \citenamefont {Gu},\ and\ \citenamefont {Wen}}]{Yang:2017hj}%
  \BibitemOpen
  \bibfield  {author} {\bibinfo {author} {\bibfnamefont {Shuo}\ \bibnamefont
  {Yang}}, \bibinfo {author} {\bibfnamefont {Zheng-Cheng}\ \bibnamefont {Gu}},
  \ and\ \bibinfo {author} {\bibfnamefont {Xiao-Gang}\ \bibnamefont {Wen}},\
  }\bibfield  {title} {\enquote {\bibinfo {title} {{Loop Optimization for
  Tensor Network Renormalization}},}\ }\href {\doibase
  10.1103/PhysRevLett.118.110504} {\bibfield  {journal} {\bibinfo  {journal}
  {Physical Review Letters}\ }\textbf {\bibinfo {volume} {118}},\ \bibinfo
  {pages} {110504} (\bibinfo {year} {2017})}\BibitemShut {NoStop}%
\bibitem [{\citenamefont {Hauru}\ \emph {et~al.}(2018)\citenamefont {Hauru},
  \citenamefont {Delcamp},\ and\ \citenamefont {Mizera}}]{Hauru:2018ij}%
  \BibitemOpen
  \bibfield  {author} {\bibinfo {author} {\bibfnamefont {Markus}\ \bibnamefont
  {Hauru}}, \bibinfo {author} {\bibfnamefont {Clement}\ \bibnamefont
  {Delcamp}}, \ and\ \bibinfo {author} {\bibfnamefont {Sebastian}\ \bibnamefont
  {Mizera}},\ }\bibfield  {title} {\enquote {\bibinfo {title} {{Renormalization
  of tensor networks using graph-independent local truncations}},}\ }\href
  {\doibase 10.1103/PhysRevB.97.045111} {\bibfield  {journal} {\bibinfo
  {journal} {Physical Review B}\ }\textbf {\bibinfo {volume} {97}},\ \bibinfo
  {pages} {045111} (\bibinfo {year} {2018})}\BibitemShut {NoStop}%
\bibitem [{\citenamefont {Harada}(2018)}]{Harada:2018cj}%
  \BibitemOpen
  \bibfield  {author} {\bibinfo {author} {\bibfnamefont {Kenji}\ \bibnamefont
  {Harada}},\ }\bibfield  {title} {\enquote {\bibinfo {title} {{Entanglement
  branching operator}},}\ }\href {\doibase 10.1103/PhysRevB.97.045124}
  {\bibfield  {journal} {\bibinfo  {journal} {Physical Review B}\ }\textbf
  {\bibinfo {volume} {97}},\ \bibinfo {pages} {045124} (\bibinfo {year}
  {2018})}\BibitemShut {NoStop}%
\end{thebibliography}%

\end{document}